\documentclass[aps,prstab,twocolumn,showpacs,preprintnumbers,floatfix]{revtex4}
\usepackage{graphicx}
\usepackage{amsmath}  
\usepackage{amssymb}  
\usepackage{mathrsfs}    
\usepackage{euscript}     
\usepackage{bm}		
\usepackage{verbatim}     
\def\be{\begin{equation}}      
\def\ee{\end{equation}}

\def\beu{\begin{equation*}}   
\def\eeu{\end{equation*}}

\def\bse{\begin{subequations}}  
\def\ese{\end{subequations}}

\providecommand{\abs}[1]{\left\lvert#1\right\rvert}   

\providecommand{\bv}[1]{\boldsymbol{#1}}        
\providecommand{\unitvec}[1]{\hat{\boldsymbol{#1}}} 

\begin{document}

\title{Time dependence of Bragg forward scattering and self-seeding of hard x-ray free-electron lasers}

\author{R.~R.~Lindberg}\email{lindberg@aps.anl.gov}
\author{Yu.~V.~Shvyd'ko}\email{shvydko@aps.anl.gov}
\affiliation{Advanced Photon Source, Argonne National Laboratory, Argonne, IL 60439, USA}

\begin{abstract}
Free-electron lasers (FELs) can now generate temporally short, high power x-ray pulses of unprecedented brightness, even though their longitudinal coherence is relatively poor.  The longitudinal coherence can be potentially improved by employing narrow bandwidth x-ray crystal optics, in which case one must also understand how the crystal affects the field profile in time and space.  We frame the dynamical theory of x-ray diffraction as a set of coupled waves in order to derive analytic expressions for the spatiotemporal response of Bragg scattering from temporally short incident pulses.  We compute the profiles of both the reflected and forward scattered x-ray pulses, showing that the time delay of the wave $\tau$ is linked to its transverse spatial shift $\Delta x$ through the simple relationship $\Delta x = c\tau \cot\theta$, where $\theta$ is the grazing angle of incidence to the diffracting planes.  Finally, we apply our findings to obtain an analytic description of Bragg forward scattering relevant to monochromatically seed hard x-ray FELs.
\end{abstract}

\date{\today}
\maketitle

Free-electron lasers (FELs) based on self-amplified spontaneous emission (SASE)  \cite{KondratenkoSaldin80, BonifacioPellegriniNarducci84} are a new source of x-rays whose brightness is many orders of magnitude larger than those of more traditional third-generation synchrotrons.  Nevertheless, because SASE radiation is initialized (or seeded) by the shot noise of the electron beam, it comprises many temporal/longitudinal modes \cite{Kim:1986}.  The bandwidth of this chaotic light is limited to that of the FEL process, which for high-gain FELs means that the normalized frequency bandwidth $\Delta\omega/\omega \sim \rho$, where $\rho$ is the FEL Pierce parameter that is typically between $10^{-4}$ and $10^{-3}$ for hard x-ray FELs.

While SASE FELs are limited to $\Delta\omega/\omega \sim \rho$, one can decrease the output bandwidth and increase the longitudinal coherence by seeding the FEL with coherent radiation from an external source.  For hard x-rays where an external source is not available, it was proposed that one could improve the longitudinal coherence through self-seeding -- by putting the SASE light through an x-ray monochromator one can obtain a narrow bandwidth source with which to generate longitudinally coherent FEL radiation.  The primary difficulty with the scheme proposed in 
\cite{Saldin_etal:2001} was that the monochromator delayed the radiation by several picoseconds, which in turn required a long ($\sim 40$ m) magnetic chicane to delay the electron beam by the same amount.  One way of alleviating this is to use two separate electron bunches 
\cite{Ding_etal:2010}.

Recently, an alternative seeding method was proposed \cite{Geloni_etal:2010a, Geloni_etal:2011} that takes advantage of the time response of Bragg forward scattering to generate the seed.  In this scheme, dubbed the ``wake monochromator,'' a single Bragg crystal is employed in transmission to generate a delayed, monochromatic signal (the wake) that is then used to seed the FEL.  Since the wake follows the SASE pulse by $\sim$10 fs, a modest chicane is sufficient to induce the required delay in the electron beam.  The physics discussed in \cite{Geloni_etal:2010a} was in terms of a band-stop filter whose sharp spectral feature naturally gave rise to a long, monochromatic temporal signal that was delayed due to causality.  Here, we give a more thorough analysis of the process, obtaining analytic expressions for the spatiotemporal shape and power of the monochromatic wake generated by an incident SASE pulse.

The temporal dependence of x-ray Bragg diffraction in crystals has been studied previously in several publications \cite{SZM:2001a, SZM:2001b, Graeff:2002, MalgrangeGraeff:2003, Graeff:2004, Shvydko:2004, Bushuev:2008} by Fourier transforming the frequency domain Bragg diffraction amplitudes, which are well-known from the classical dynamical theory of x-ray diffraction \cite{Zachariasen:1945, Laue:1960, BattermanCole:1964, Pinsker:1978, Authier:2001}.  Here, our focus is on the spatiotemporal dynamics that play an essential role in the efficacy of the FEL self-seeding.  For this reason, the problem is solved directly in time and space using a system of space-time coupled wave equations.

We begin Sec.~\ref{sec:BraggScatter} by deriving the coupled wave system relevant to symmetric Bragg scattering, and subsequently solve for the field profiles in both reflection and transmission.  We show that the temporal profile consists of a sequence of power maxima whose location depends on the crystal extinction length $\Lambda$ and grazing incidence angle $\theta$ in reflection, while in transmission the time delay also varies inversely with the crystal thickness $d$.  In addition, the profiles are displaced along the transverse direction $x_o$ by an amount proportional to the delay $\tau$: $\Delta x_o = c\tau\cot\theta$.  Finally, we use some well-known FEL physics to apply the theory of forward scattering to the FEL self-seeding scheme, obtaining analytic results associated with the wake monochromator of Ref.~\cite{Geloni_etal:2010a}.

\section{Bragg scattering in the time domain}	\label{sec:BraggScatter}

\begin{figure}
\centering
\includegraphics[scale=1]{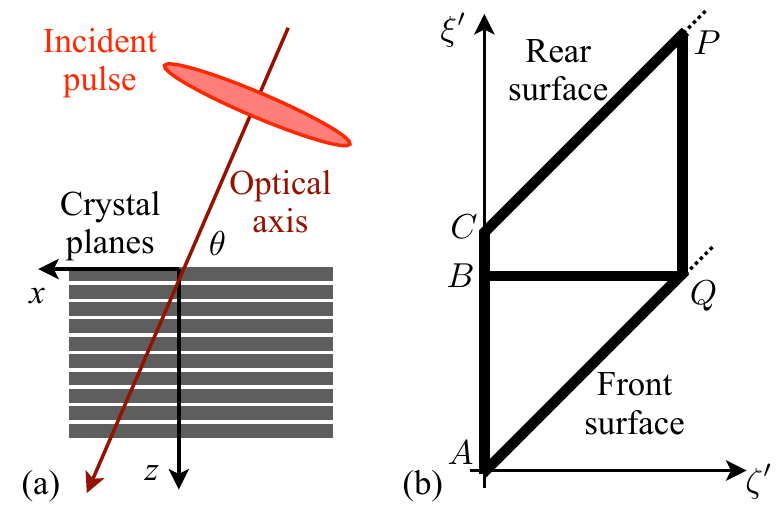}
\caption{Geometry of Bragg scattering.  (a) shows that the incident pulse travels along the optical axis that makes an angle $\theta$ with respect to the crystal surface.  (b) plots the geometry in the characteristic coordinate plane $\zeta'$-$\xi'$, with the crystal entrance and rear surfaces located at $\zeta' =\xi'$ (the line $\overline{AQ}$) and $\zeta' =\xi'-d$ (the line $\overline{CP}$), respectively.  We also include the integration contours $ABQ$ and $ACPQ$ that are used to obtain the solution for the reflected and transmitted waves.}	\label{fig:1}
\end{figure}

We are predominantly interested in understanding the basic physics and deriving a few simple analytic expressions for the time response of Bragg forward scattering from temporally short incident and laterally confined pulses.  To simplify the subsequent analysis and avoid notational complications, we assume that the crystal is symmetric, meaning that the crystal planes responsible for Bragg scattering are parallel to the crystal surface.  This restriction is most relevant to the FEL monochromator we subsequently study.  We take the optical axis of the incident radiation to make an angle $\theta$ from the surface as shown in Fig.~\ref{fig:1}(a).  Maxwell's wave equation for the electric field $\bv{E}$ in the crystal is
\be
\begin{split}
  \left[ \frac{1}{c^2} \frac{\partial^2}{\partial t^2} - \bv{\nabla}^2 \right] \bv{E}(\bv{r}, t) = -\frac{4\pi}{c^2}
  	\frac{\partial \bv{J}}{\partial t} - \bv{\nabla}(\bv{\nabla} \cdot \bv{E}),&	\label{eqn:MaxwellWave}
\end{split}
\ee
where $\bv{J}$ is the current density induced in the crystal by the radiation and $c$ is the speed of light.  The linear response of the medium is given by the crystal polarizability $\chi(\bv{r})$.  Denoting the polarization components by a superscript $s$, in the linear approximation for x-ray energies far above any atomic resonances we have
\begin{align}
  J^s(\bv{r}, t) &= -\int \! d\omega \; e^{-i\omega t}\frac{i\omega}{4\pi} \sum_{s'}\chi^{ss'}(\bv{r})
  	E^{s'}(\bv{r}, \omega) \nonumber  \\
  &= \frac{1}{4\pi} \sum_{h,s'} \chi_h^{ss'}e^{i\bv{h}\cdot\bv{r}} \frac{\partial}{\partial t}E^{s'}(\bv{r}, t),
  	\label{eqn:Current}
\end{align}
where in the second line we have used the periodicity of the crystal to expand the polarizability $\chi$ as a Fourier series.  We decompose the electric field vector as a sum of two orthogonal polarizations, $\bv{E} = E^{\sigma} \unitvec{e}_\sigma + E^{\pi} \unitvec{e}_\pi$, with  $\unitvec{e}_\sigma$ parallel and $\unitvec{e}_\pi$ perpendicular to the scattering plane.  In this case, the non-diagonal components (namely, $\chi_h^{ss'}$ for $s' \ne s$) vanish and the polarization components decouple.  
For simplicity, we assume $\sigma$-polarization and subsequently drop the polarization dependence.

The electronic polarizability in \eqref{eqn:Current} strongly couples electromagnetic waves whose wave-vectors differ by $\bv{h}$, which results in strong Bragg reflection.  We assume that this condition is satisfied for one reciprocal lattice vector, which in the symmetric geometry implies that $\bv{h} = -h\unitvec{z}$.   Since the $\chi_{\vphantom{\bar{h}}h}$ are assumed to be time independent, it is natural to take a Fourier transform with respect to $t$; however, our goal is to compute the temporal response, and we find it more convenient to remain in the time domain.  Instead, we introduce the slowly varying eikonal field amplitudes associated with two interacting waves, writing the electromagnetic wave as
\be
\begin{split}
  E(\bv{x}, z, t) &= e^{-ick_0 t} e^{i k_0[z \sin\theta +  x\cos\theta]} \\
  &\phantom{=}\times\left[E_0(\bv{x}, z, t) + e^{-i h z} E_h(\bv{x}, z, t) \right],
  	\label{eqn:Eikonal}
\end{split}
\ee
where $ck_0$ is the carrier frequency and $(k_0\sin\theta, 0, k_0\cos\theta)$ is the reference wave-vector.  We assume that the field envelope functions have slow spatiotemporal variations with respect to the carrier frequency $ck_0$, with
\begin{align}
  \abs{\frac{\partial}{\partial t} \ln E_{0,h}} &\ll ck_0, &  
  	\abs{\frac{\partial}{\partial z} \ln E_{0,h}} &\ll k_0 \sin\theta.	\label{eqn:SlowlyVary}
\end{align}
The reference carrier frequency $ck_0$ can be freely chosen provided that \eqref{eqn:SlowlyVary} hold, with any additional slow temporal dependence being accounted for by $E_{0,h}$; 
we choose $k_0$ to satisfy Bragg's condition: $k_0\sin\theta = h/2$.  

We have oriented our axes such that the scattering plane defined by the reciprocal lattice vector and the optical axis lie in the $(x,z)$ plane, and we assume that the crystal is uniform in the $(x,y)$ plane.  Thus, the fundamental field solutions are plane waves in the directions parallel to the crystal surface.  For simplicity we will neglect the trivial $y$-dependence, with the general electric field $E$ being written as the superposition
\be
  E_{0,h}(x, z, t) = \int \! dq \; e^{i q x} \mathcal{E}_{0,h}(q, z, t).
  	\label{eqn:TransFT}
\ee

We now use the slowly-varying assumption \eqref{eqn:SlowlyVary} to reduce the second order wave equation \eqref{eqn:MaxwellWave} to two first-order equations for the field amplitudes $\mathcal{E}_0$ and $\mathcal{E}_h$, an approach that was first developed in this context by Takagi \cite{Takagi:1962, Takagi:1969}.  Since we assume that the crystal lattice is perfect, we find it convenient to use $z$ and $t$ as the relevant coordinates rather than the standard spatial coordinates along the wave-vectors (see, e.g., \cite{Pinsker:1978, Authier:2001, AfanasevKohn:1971, ChukhovskiiForster:1995}); thus, our approach is similar to those of \cite{KohnShvydko:1995, WarkLee:1999}.  We insert the forms \eqref{eqn:TransFT} and \eqref{eqn:Eikonal} into the wave equation \eqref{eqn:MaxwellWave}, and drop the term $\sim \bv{\nabla}(\bv{\nabla}\cdot\bv{E})$ since the field remains approximately transverse in the crystal.  By neglecting higher-order derivatives of the field amplitudes and matching fast phases, we obtain the following set of slowly-varying equations:
\begin{align}
  \left[ \frac{\partial}{\partial ct} + \sin\theta \frac{\partial}{\partial z}
  	+ ik_0\left(\tilde{\alpha}_0 - \frac{\chi_0}{2}\right) \right]
	\mathcal{E}_0 &= \frac{ik_0\chi_{\bar{h}}}{2} \mathcal{E}_h  \label{eqn:E0}  \\
  \left[ \frac{\partial}{\partial ct} - \sin\theta \frac{\partial}{\partial z}
  	+ ik_0\left(\tilde{\alpha}_h - \frac{\chi_0}{2}\right) \right] 
	\mathcal{E}_h &= \frac{ik_0\chi_{\vphantom{\bar{h}}h}}{2} \mathcal{E}_0, 	\label{eqn:EH}
\end{align}
where for convenience we define $\chi_{\bar{h}} \equiv \chi_{\vphantom{\bar{h}}-h}$.  We have attached tildes to $\tilde{\alpha}_{0,h}$ because while they appear similar to the usual deviation from Bragg's condition $\alpha_{0,h}$, they have a slightly different definition and interpretation in the present context.  $\tilde{\alpha}_0$ represents the incidence wave-vector's difference from the vacuum condition due to $q \ne 0$, while $\tilde{\alpha}_h$ is the deviation of the reflected carrier wave from Bragg's condition, with
\begin{align}
  \tilde{\alpha}_0 &\equiv \frac{1}{2k_0^2}\big[ k_0^2 - k_0^2\sin^2\theta - (k_0\cos\theta + q)^2 \big]  \\
  \tilde{\alpha}_h &\equiv \frac{1}{2k_0^2}\big[ k_0^2 - (k_0\sin\theta - h)^2 - (k_0\cos\theta + q)^2 \big];
\end{align}
our choice $k_0\sin\theta = h/2$ implies that
\be
  \tilde{\alpha}_0 = \tilde{\alpha}_h = \frac{q}{k_0} \cos\theta + \frac{q^2}{2k_0^2} \equiv \tilde{\alpha}.
\ee
Note that the central wavenumber $k_0$ and the crystal parameters $\chi_0$, $\chi_{\vphantom{\bar{h}} h}$, and $\chi_{\bar{h}}$ are all functions of angle $\theta$.  To solve \eqref{eqn:E0}-\eqref{eqn:EH}, we introduce the characteristic coordinates
\begin{align}
  \zeta &\equiv \tfrac{1}{2} \left( ct \sin\theta - z\right),  &  \xi &\equiv \tfrac{1}{2} \left( ct \sin\theta + z\right),
\end{align}
and the reduced field amplitudes $\mathcal{A}_0$ and $\mathcal{A}_h$ via
\begin{align}
  \mathcal{E}_0 &\equiv e^{-ik_0(\tilde{\alpha} - \chi_0/2)(\zeta + \xi)/\sin\theta} \mathcal{A}_0
  		\label{eqn:reducedDef_0}  \\
  \mathcal{E}_h &\equiv e^{-ik_0(\tilde{\alpha} - \chi_0/2)(\zeta + \xi)/\sin\theta} \mathcal{A}_h.
	  	\label{eqn:reducedDef_h}
\end{align}
Then, the coupled wave system \eqref{eqn:E0}-\eqref{eqn:EH} reduces to
\begin{align}
  \frac{\partial\mathcal{A}_0}{\partial\xi} &= \frac{ik_0\chi_{\bar{h}}}{2\sin\theta}\mathcal{A}_h,  &
  	\frac{\partial\mathcal{A}_h}{\partial\zeta} &= 
	\frac{ik_0\chi_{\vphantom{\bar{h}} h}}{2\sin\theta}\mathcal{A}_0.	\label{eqn:ReducedSystem}
\end{align}
Here, we see that if the coupling $\chi_{\vphantom{\bar{h}}h} \rightarrow 0$, the forward-going wave $\mathcal{A}_0$ is a function of $\zeta$ only, while the reflected wave $\mathcal{A}_h$ is a fixed function of $\xi$.  In the crystal, the two waves interact as shown, with each obeying the associated second-order equation
\be
  \frac{\partial^2}{\partial\xi\partial\zeta}\mathcal{A} = 
  	-\frac{k_0^2\chi_{\bar{h}}\chi_{\vphantom{\bar{h}}h}}{4\sin^2\theta}\mathcal{A} \equiv 
	-\frac{\pi^2}{\Lambda^2}\mathcal{A},
\ee
where the extinction length 
$\Lambda \equiv (2\pi/k_0) \sin\theta/\sqrt{\chi_{\vphantom{\bar{h}}h}\chi_{\bar{h}}}$ is approximately independent of the angle $\theta$ for any given reflection [this can be seen by using $\chi_{\vphantom{\bar{h}}h} \approx \Re(\chi_{\vphantom{\bar{h}}h}) \sim \sin^2\theta$ and $k_0 = h/(2\sin\theta)$].

The linear system \eqref{eqn:ReducedSystem} can be solved for specified boundary conditions using the Riemann method as was done in \cite{AfanasevKohn:1971}, and which is reviewed in, e.g., \cite{Authier:2001} and in the Appendix.  For Bragg scattering, the reduced fields satisfy \eqref{eqn:ReducedSystem} subject to the constraints that the forward-going wave $\mathcal{A}_0$ is a prescribed function along the front crystal surface at $z=0$ while the reflected wave vanishes along the rear surface defined by $z=d$.  In terms of the characteristic coordinates shown in Fig.~\ref{fig:1}(b), we have
\begin{align}
  \mathcal{A}_0 \big\rvert_{\overline{AQ}} &= \mathcal{A}_{\text{inc}},  &
  	\mathcal{A}_h \big\rvert_{\overline{CP}} &= 0.	\label{eqn:BraggBCs}
\end{align}

In the following two sections we solve the system \eqref{eqn:ReducedSystem} subject to \eqref{eqn:BraggBCs} assuming a Gaussian incident pulse that is temporally short and confined laterally in space.  We will find relatively simple, approximate analytic expressions for the electric field when we can neglect multiple scattering of the waves off the crystal surfaces, meaning that we only consider the time interval following the main pulse that is smaller than $(2d/c)/\sin\theta$.  For a crystal of thickness $d \simeq 0.1$ mm this time interval is $\simeq (300/\sin\theta)$ fs.  The field at longer times can be built up by inserting these solutions into the Riemann integrals in an iterative manner \cite{Uragami:1969, AfanasevKohn:1971}, but this is beyond the scope of the present study.

\subsection{Bragg diffraction: the reflected wave}

In the Appendix we use Riemann's method to show that on the front crystal surface the reflected field is given by
\be
  \mathcal{A}_h(Q) = \int\limits_A^Q \! d\zeta' \; \frac{ik_0\chi_{\vphantom{\bar{h}}h}}{2\sin\theta}
  	\mathcal{A}_{\text{inc}}(\zeta',\zeta') R_h(\zeta, \xi; \zeta', \zeta'),	\label{eqn:Refl_IntSol}
\ee
where $\mathcal{A}_{\text{inc}}$ is the initially incident wave and the Riemann function $R_h$ for the reflected wave satisfies the adjoint equation associated with the (formally self-adjoint) system \eqref{eqn:ReducedSystem}
\be
  \frac{\partial^2}{\partial\xi'\partial\zeta'}R_h(\zeta, \xi; \zeta', \xi') = -\frac{\pi^2}{\Lambda^2}
  	R_h(\zeta, \xi; \zeta', \xi'), 	\label{eqn:Adjoint}
\ee
along with the auxiliary conditions
\begin{align}
  \left. \frac{\partial R_h}{\partial\xi'}\right\rvert_{\overline{AQ}} &= 0,  &
  	\left. \frac{\partial R_h}{\partial\zeta'}\right\rvert_{\overline{BQ}} &= 0,  &  R_h(Q) &= 1.
		\label{eqn:ReflectedAux}
\end{align}
The Riemann function satisfying \eqref{eqn:Adjoint}-\eqref{eqn:ReflectedAux} is given by 
\eqref{eqn:RiemannReflAppend} \cite{AfanasevKohn:1971}; for the solution \eqref{eqn:Refl_IntSol} we need $R_h$ along the front crystal surface defined by the line $\overline{AQ}$; here, we have $\xi'=\zeta'$ with point $Q$ located at $\xi=\zeta$, for which
\be
  R_h(\zeta, \zeta; \zeta', \zeta') = 2\frac{J_1[2\pi(\zeta-\zeta')/\Lambda]}{2\pi(\zeta-\zeta')/\Lambda}.
  	\label{eqn:ReflRiemannLine}
\ee

\begin{figure}
\centering
\includegraphics[scale=1]{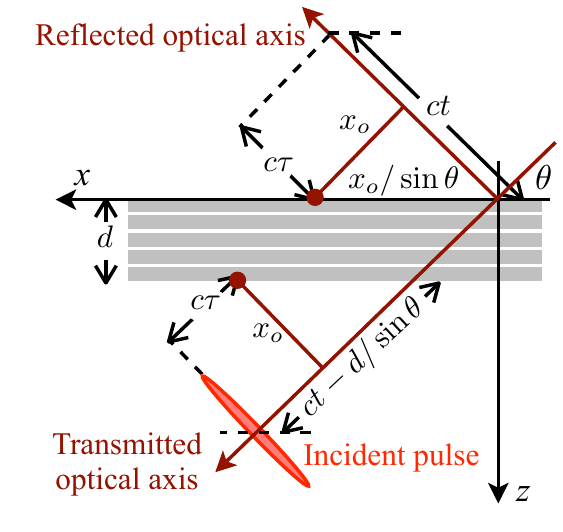}
\caption{Geometric relationship between the crystal coordinates $(x, ct)$ and the optical axis coordinates $(x_o, c\tau)$ for both the reflected (top) and transmitted (bottom) wave.}	\label{fig:2}
\end{figure}

The physical electric field amplitude of the reflected wave can be obtained by applying the field definitions \eqref{eqn:reducedDef_0}-\eqref{eqn:reducedDef_h} to the solution \eqref{eqn:Refl_IntSol} once we specify the incident field.  We will be interested in the response from a temporally short incident wave directed along the optical axis shown in Fig.~\ref{fig:1}(a).  We model the incident wave by a Gaussian field both longitudinally and transversely that propagates along the optical axis defined by $\unitvec{z} \sin\theta + \unitvec{x} \cos\theta$:
\beu
\begin{split}
  E_{\text{inc}} &= \frac{1}{\sqrt{4\pi}\sigma_\tau}\exp\!\left\{
  	-\frac{1}{4\sigma_\tau^2}[ct - (z\sin\theta + x\cos\theta)]^2\right\}  \\
  &\phantom{=|\sqrt{4\pi}\sigma_\tau mm}
  	\times \exp\!\left[ -\frac{1}{4\sigma_x^2}(x\sin\theta - z\cos\theta)^2\right].
\end{split}
\eeu
If the incident pulse length $\sigma_\tau$ is much shorter than the penetration length, we can approximate the temporal profile as a delta-function; taking the limit $\sigma_\tau \rightarrow 0$ and using the definitions \eqref{eqn:TransFT}, \eqref{eqn:reducedDef_0}, and \eqref{eqn:reducedDef_h}, the incident field along the crystal surface $z=0$ in terms of the characteristic coordinates is
\begin{align}
  \mathcal{A}_{\text{inc}} &=  \frac{1}{2\pi} \int \! dx' \; e^{-iqx'} e^{ik_0(2\tilde{\alpha}-\chi_0)\zeta'/\sin\theta}
  	E_{\text{inc}}	\nonumber  \\
\begin{split}
  &\rightarrow \frac{1}{2\pi} \int \! dx' \; e^{-iqx'} e^{ik_0(2\tilde{\alpha} - \chi_0)\zeta'/\sin\theta} 
  	e^{-x'^2\sin^2\theta/4\sigma_x^2}   \\
  &\phantom{=mmmmm} \times \frac{\sin\theta}{2}\delta(\zeta' - x'\sin\theta\cos\theta/2).	\label{eqn:IC}
\end{split}
\end{align}
Upon inserting the initial condition \eqref{eqn:IC} into the solution for $\mathcal{A}_h$ \eqref{eqn:Refl_IntSol}, the integral over $\zeta'$ can be trivially performed.  To get the physical reflected field $E_h$ requires the inverse Fourier transform as indicated by \eqref{eqn:TransFT}.  The transform with respect to $q$ is a Gaussian integral that can be taken analytically, so that
\be
\begin{split}
  E_h &= \frac{ik_0\chi_{\vphantom{\bar{h}}h}}{2\sqrt{2\pi}} \int \! dx' \; \sqrt{\frac{-ik_0}{ct-x'\cos\theta}}  \\
  &\phantom{=|mmm}  \times e^{i\chi_0 k_0(ct-x'\cos\theta)/2}e^{-{x'}^2\sin^2\theta^2/4\sigma_x^2} \\
  &\phantom{=| mmm} \times \exp\!\left[ 
  	\frac{ik_0(x'\sin^2\theta - x + ct\cos\theta)^2}{2(ct - x'\cos\theta)}\right]  \\
  &\hspace{1 in} \times \frac{J_1[\pi \sin\theta(ct - x'\cos\theta)/\Lambda]}
	{\pi \sin\theta(ct - x'\cos\theta)/\Lambda}.	\label{eqn:EreflSolved}
\end{split}
\ee
Equation \eqref{eqn:EreflSolved} expresses the reflected wave from the temporally short incident field \eqref{eqn:IC}.  Before attacking the integral, we must first relate the coordinates $(x,t)$ to those along the new optical axis defined by the reflected wave.  As depicted in Fig.~\ref{fig:2}, these comprise the coordinate transverse to the optical axis $x_o$ and the orthogonal time delay $\tau$ defined at $z=0$.  Note that the time difference is defined such that lines of constant $\tau$ are parallel to the $x_o$-axis.  Since the time $t$ is measured with respect to lines of constant $z$, the reflected wave optical axis coordinates are related to $(x, t)$ via $x = x_o/\sin\theta$ and $ct = c\tau + x_o/\tan\theta$ as shown in Fig.~\ref{fig:2}.

To make further analytic progress, we can approximate this integral using the method of stationary phase.  Computation of the stationary points is greatly simplified if we assume that $\abs{\chi_0} \ll \sin^4\theta$ and that the transverse size $\sigma_x$ is sufficiently large.  Specifically, we assume that
\begin{align}
  k_0 \sigma_x^2 &\gg \frac{c\tau}{\sin^2\theta}  & \frac{1}{k_0\sigma_x} \ll \tan\theta \sin^2\theta.
\end{align}
Physically, the first condition means that we consider distances behind the incident pulse $c\tau$  that are much less than the Rayleigh range associated with the beam spot size projected onto the crystal surface, while the second condition is equivalent to requiring the beam angular divergence is much less than $\sin^3\theta$; these conditions are typically well-satisfied for all but extreme grazing angles.  Under these assumptions, the stationary point is
\be
  x_s' = \frac{x- ct\cos\theta}{\sin^2\theta} = \frac{x_o}{\sin\theta} - \frac{c\tau \cos\theta}{\sin^2\theta},
  	\label{eqn:stationaryPt}
\ee
and the reflected wave is given by
\be
\begin{split}
  E_h &= \frac{ik_0\chi_{\vphantom{\bar{h}}h}}{2\sin^2\theta}
  	e^{i\chi_0 ck_0 \tau/2\sin^2\theta}
  	\frac{J_1[\pi c\tau/(\Lambda \sin\theta)]}{\pi c\tau/(\Lambda \sin\theta)}  \\
  &\phantom{=|mmm} \times \exp\!\left\{ -\frac{1}{4\sigma_x^2}
  	\left[(x_\text{o} - c\tau\cot \theta)^2\right] \right\}. 	\label{eqn:ReflectedFinalSol}
\end{split}
\ee

\begin{figure*}
\centering
\includegraphics[scale=1]{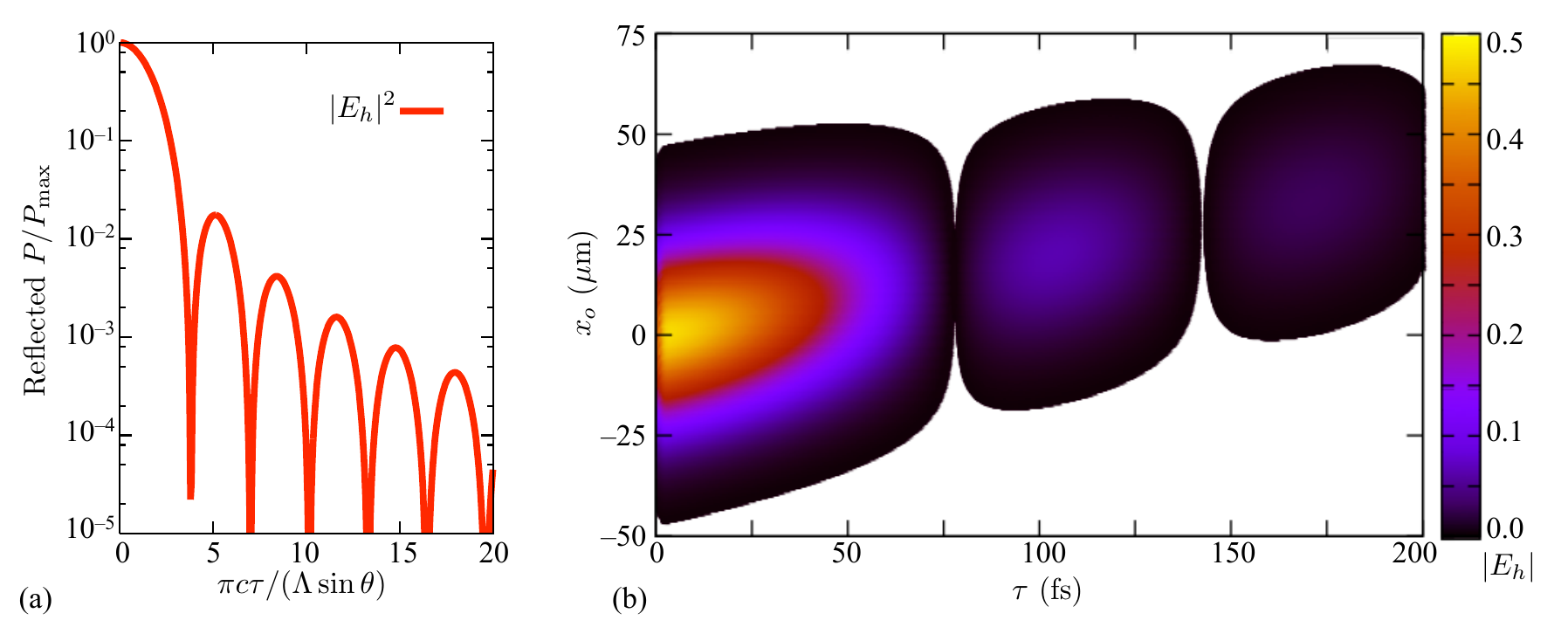}
\caption{(a) The temporal power profile $\sim \abs{J_1(y)/y}^2$ of the Bragg reflected wave as a function of argument $\pi c\tau/(\Lambda \sin\theta)$.  (b) Reflected field magnitude $\abs{E_h}$ from the (004) Bragg reflection of diamond at an incidence angle $\theta = 56.86^\circ$ when the incident wave has $\sigma_x = 10 \; \mu\text{m}$.  The wave is displaced transversely to the optical axis by an amount proportional to the time elapsed, with $x_o = c\tau \cot\theta$.
}	\label{fig:3}
\end{figure*}

From \eqref{eqn:ReflectedFinalSol}, the temporal profile of the reflected wave oscillates over the time-scale $(\Lambda/c)\sin\theta$ according to the Bessel function $J_1$; we graph the sequence of power peaks associated with $\abs{J_1(y)/y}^2$ in Fig.~\ref{fig:3}(a).  Additionally, the field profile is laterally displaced along $x_o$ as the time delay increases according to $x_o = c\tau\cot \theta$.  We plot a specific example of this behavior in Fig.~\ref{fig:3}(b) for the C(004) reflection at $56.85^\circ$, for which $\Lambda \approx 22.8 \; \mu\text{m}$ and $\lambda_0 \approx 1.5$ \AA.  Figure \ref{fig:3}(b) plots the reflected amplitude $\abs{E_h}$ to more clearly identify the trailing pulses, which are displaced as predicted by Eq.~\eqref{eqn:ReflectedFinalSol}.  In the next section we apply a similar analysis to forward scattering.  We will show that while the behavior of the transverse envelope is quite similar, the temporal profile of Bragg forward scattering is distinct, in that it depends importantly on the crystal thickness $d$.

\subsection{Bragg forward scattering}

Having found the spatiotemporal dependence of the reflected wave, we now turn to obtaining the transmitted wave $E_0$ at the rear surface of the crystal.  The solution can again be determined by Riemann's method; restricting ourselves to $d/\sin\theta \le ct \le 3d/\sin\theta$, in the Appendix we show that the transmitted wave is given by
\be
\begin{split}
  \mathcal{A}_0\big\rvert_P &= R_0 \mathcal{A}_0\big\rvert_Q - \int\limits_A^Q \! d\zeta' \;
	R_0(\zeta,\xi;\zeta',\zeta')\frac{ik_0\chi_{\bar{h}}}{2\sin\theta}\mathcal{A}_h  \\
  &\phantom{=} -\int\limits_A^Q \! d\zeta' \, \left.\mathcal{A}_0(\zeta',\zeta')\frac{\partial}{\partial\zeta'}
  	R_0(\zeta,\xi;\xi',\zeta')\right\rvert_{\xi'=\zeta'}, 	\label{eqn:TransSol}
\end{split}
\ee
where the Riemann function that satisfies the relevant boundary conditions \eqref{eqn:TransAux} and the adjoint equation \eqref{eqn:Adjoint} is \cite{AfanasevKohn:1971}
\be
\begin{split}
  R_0(\zeta, \xi; \zeta', \xi') = J_0\!\left[2\pi \sqrt{(\zeta-\zeta')(\xi-\xi')}/\Lambda \right]  \hspace{.25 in} \\
  + \frac{\zeta-\zeta'}{\xi-\xi'}J_2\!\left[2\pi \sqrt{(\zeta-\zeta')(\xi-\xi')}/\Lambda \right].
\end{split}
\ee
Along the exit crystal surface $\overline{CP}$ we have $\xi = \zeta + d$, so that the second integrand in \eqref{eqn:TransSol} involves
\be
  \frac{\partial R_0}{\partial\zeta'} = \frac{2\pi^2 d}{\Lambda^2} \,
  	\frac{J_1\!\left[2\pi \sqrt{(\zeta-\zeta')(\zeta+d-\zeta')}/\Lambda \right]}
	{2\pi \sqrt{(\zeta-\zeta')(\zeta+d-\zeta')}/\Lambda}.		\label{eqn:TransRiemann}
\ee

The first term in \eqref{eqn:TransSol} is merely the initial condition evolved along the characteristics, which is not the focus of this study, and we will henceforth drop it from our expressions.  Of the remaining two terms, the one on the second line in \eqref{eqn:TransSol} is generated by the reflected wave that is directly excited by the incident pulse, which can be shown to dominate the second term from the first line in the following way.  Since $\mathcal{A}_0 \propto \delta(\zeta' - x\cos\theta\sin\theta/2)$, upon integration the second line scales as the product of $1/\Lambda^2$ and the x-ray path length through the crystal $d/\sin\theta$.  Using the expressions \eqref{eqn:Refl_IntSol} and \eqref{eqn:IC} for $\mathcal{A}_h$, it is easily shown that the integral on the first line in \eqref{eqn:TransSol} scales as $1/\Lambda^2$ times a highly oscillatory function integrated over the time $ct$; for the times $ct \lesssim d/\sin\theta$ that we are considering, the first line of \eqref{eqn:TransSol} is therefore negligible with respect to that of the second line.

\begin{widetext}
Thus, the transmitted wave of interest is given by the second line in \eqref{eqn:TransSol}.  An expression for the transmitted electric field envelope is obtained by proceeding as we did for the reflected wave; we use the definitions of the reduced fields \eqref{eqn:reducedDef_0}-\eqref{eqn:reducedDef_h}, insert the function \eqref{eqn:TransRiemann}, and use the initially short incident pulse \eqref{eqn:IC}.  The integration over $\zeta'$ is then trivial, and we find
\be
  \mathcal{E}_0\big{\rvert}_P = -\frac{\sin\theta}{4 \pi} 
  	\int \! dx' \; e^{-iqx'} e^{-x'^2\sin^2\theta/4\sigma_x^2}
	e^{-ik_0(\tilde{\alpha} - \chi_0/2)(2\zeta/\sin\theta + d/\sin\theta - x'\cos\theta)} \left. 
	\frac{\partial R_0}{\partial\zeta'} \right\rvert_{\zeta'=(x'/2)\cos\theta\sin\theta}.
		\label{eqn:TransSol2}
\ee
While this expression is difficult to interpret, we can make further progress by considering the field $E_0$ in physical space.  We apply $\int dq \, e^{iqx}$ to both sides of \eqref{eqn:TransSol2}; again, the integral over $q$ is a Gaussian one that can be taken analytically, leading to
\be
\begin{split}
  E_0\big{\rvert}_P &= -\frac{\pi^2d\sin\theta}{\Lambda^2}  \int \! dx' \; 
  	\frac{\sqrt{-ik_0} \; e^{-x'^2\sin^2\theta/4\sigma_x^2}}
	{\sqrt{2\pi [(d + 2\zeta)/\sin\theta - x'\cos\theta]}}
	\exp\!\left\{\frac{ik_0[x' \sin^2\theta - x + (2\zeta + d)\cot\theta]^2}
	{2[(d + 2\zeta)/\sin\theta - x'\cos\theta)]} \right\}   \\
  & \hspace{1.23 in} \times e^{i\chi_0 k_0[(d + 2\zeta)/\sin\theta - x'\cos\theta]/2}
    	\frac{J_1\!\left[\pi\sqrt{(2\zeta - x'\cos\theta\sin\theta)(2d + 2\zeta - x'\cos\theta\sin\theta)}
	/\Lambda\right]}{\pi\sqrt{(2\zeta - x'\cos\theta\sin\theta)(2d + 2\zeta - x'\cos\theta\sin\theta)}
	/\Lambda}.	\label{eqn:TransSolNasty}
\end{split}
\ee
Finally, the expression \eqref{eqn:TransSolNasty} should be written in term of the coordinates along the optical axis illustrated in Fig.~\ref{fig:2}, where $x_\text{o}$ again labels the transverse position with respect to the transmission optical axis, while $\tau$ is the relative time along the axis.  From the Figure, we see that at the rear surface $z=d$ we have $x = (x_\text{o} + d\cos\theta)/\sin\theta$, and $2\zeta/\sin\theta = ct - d/\sin\theta = c\tau + x_\text{o}/\tan\theta$; these coordinates will be used in the final results.
\end{widetext}
While the expression \eqref{eqn:TransSolNasty} doesn't look any more appealing than \eqref{eqn:TransSol2}, we can begin to make sense of it by considering the limit of exact backscattering, i.e., $\theta \rightarrow \pi/2$, in which case the integral over $x$ is merely the Gaussian integral associated with paraxial evolution.  Furthermore, the transverse crystal coordinate equals its optical axis counterpart $x = x_\text{o}$, while $2\zeta = c\tau$.  In the limit of exact backscattering $\theta = \pi/2$, \eqref{eqn:TransSolNasty} simplifies to
\be
\begin{split}
  E_0\big{\rvert}_P = -\frac{d\pi^2}{\Lambda^2}
  	e^{i\chi_0 k_0(d + c\tau)/2} 
  	\frac{\exp\!\left[-\frac{x_o^2}{4\sigma_x^2 - i(d+c\tau)/k_0}\right]}{1-i(d+c\tau)/2k_0\sigma_x^2}\;&  \\
  \times \frac{J_1\!\left[\pi\sqrt{c\tau(2d + c\tau)}/\Lambda\right]}{\pi\sqrt{c\tau(2d + c\tau)}/\Lambda}.&
  	\label{eqn:TransSolBackscatter}
\end{split}
\ee

\begin{figure*}
\centering
\includegraphics[scale=1]{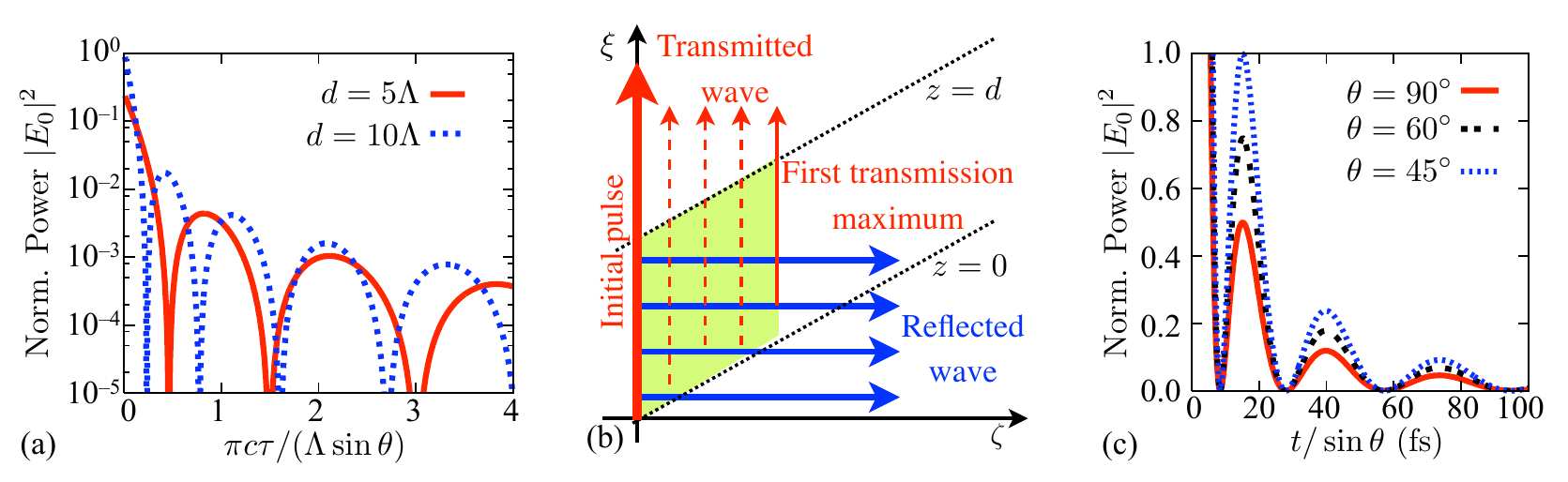}
\caption{(a) Time dependence of the Bragg forward diffraction intensity for a crystal thickness that is 5 and 10 extinction lengths.   The position of first intensity maximum scales as $1/d$ while its peak power $\propto (d/\sin\theta)^2$.  Note that in both cases there are several transmission maxima over a time interval for which Fig.~\ref{fig:3}(a) shows only one single reflection peak.  (b) Illustration of the coupled forward and reflected modes in Bragg diffraction.  The crystal surfaces are shown as dotted lines with unit slope.  The initial pulse excites a reflected wave (blue arrows) that propagates along the $\zeta$ characteristic, which in turn excites forward scattered waves (dotted red arrows) that move along $\xi$.  The first transmission maximum is determined by the coupling area shaded green, meaning that the time delay of the first maximum scales inversely with $d$.  (c) Time dependence of Bragg forward diffraction intensity, plotted as a function of $t/\sin\theta$ for the (004) reflection in diamond with $d = 0.15$ mm.  The power maxima line up on the normalized $t/\sin\theta$ scale, with the peak scaling as $(d/\sin\theta)^2$.  The central wavelength $\lambda_0 \approx (1.79 \text{ \AA})\sin\theta$.}	\label{fig:4}
\end{figure*}

In this form, the output field has a simple physical interpretation.  The first factor shows that the amplitude scales as the product of the propagation distance and square of the coupling $\pi/\Lambda$, since the incident wave must effectively be first scattered into the reflected wave and then back to the transmitted to produce the trailing pulse.  Additionally, the amplitude is altered by both the phase difference and loss due to the crystal index of refraction and by the natural transverse spreading of the pulse: in the limit $k_0\sigma_x^2 \gg (d+c\tau)$, namely, that the Rayleigh range is much greater than the total propagation distance $d + c\tau$, it yields the transverse envelope $e^{-x_o^2/4\sigma_x^2}$.  Finally, the second line gives the transmission temporal profile or the ``wake'' of the incident pulse.

We plot the temporal power profile $\sim \abs{J_1(y)/y}^2$ as a function of $\pi c\tau/\Lambda$ for two crystal thicknesses in Fig.~\ref{fig:4}(a).  Doubling the crystal thickness from $5\Lambda$ to $10\Lambda$ decreases the time between maxima while increasing the peak power of each pulse by a factor of four.  Note that the entire domain of the graph lies in the region of the first reflection maximum plotted in Fig~\ref{fig:3}(a); the characteristic time scale of the forward-scattered wave is much shorter than that of the reflected wave for $d \gtrsim \Lambda$.  Additionally, the temporal profile of forward scattering depends strongly on the crystal thickness $d$.

Equation \eqref{eqn:TransSolBackscatter} implies that the successive power maxima of the transmitted field are given by the maxima of the function $\abs{J_1(y)/y}$, which correspond to the zeroes of the Bessel function $J_2(y)$.  We denote the positions of the $\abs{J_1(y)/y}$ maxima by $\mathscr{J}_n$ for integer $n \ge 0$, with the $n=0$ peak at $\mathscr{J}_0 = 0$ and the first trailing maximum of $\abs{J_1(y)/y}$ given by $y=\mathscr{J}_1$.  Setting the argument $\pi^2c\tau(2d + c\tau)/\Lambda^2 = \mathscr{J}_n^2$, we find
\be
  \tau^{\text{max}}_n(\theta = \pi/2) = \frac{1}{c}\sqrt{d^2 + \frac{\mathscr{J}_n^2\Lambda^2}{\pi^2}}
  	- \frac{d}{c} \approx \frac{\mathscr{J}_n^2}{2\pi^2} \frac{\Lambda^2}{cd}	\label{eqn:maxima0}
\ee
if the crystal thickness is much greater than the extinction length, $d \gg \Lambda$.  The time delay of the power maxima \eqref{eqn:maxima0} are inversely proportional to the crystal thickness $d$.  This can be understood in a heuristic manner by considering Fig.~\ref{fig:4}(b).  An initially sharp incident wave $E_0$, indicated in Fig.~\ref{fig:4}(b) by the thick, red vertical arrow at $\zeta=0$, generates a reflected wave shown schematically as the blue horizontal lines directed along the $\xi$ characteristics.  As the reflected wave amplitude grows, it in turn couples to the transmitted wave (the red dotted lines), which increases its amplitude as it extracts energy from $E_h$.  In this way, energy alternates between the reflected and transmitted waves, as evidenced by the oscillatory profiles \eqref{eqn:ReflectedFinalSol} and \eqref{eqn:TransSolBackscatter}.  The position of the first transmission maximum, for example, should be given by the integrated amplitude of the reflected wave just behind the incident pulse, which is proportional to the interaction volume given by the green shaded area in Fig.~\ref{fig:4}(b).  If we denote the first transmission maximum as $\tau_1^{\text{max}}$, we have $d \tau_1^{\text{max}} = \text{constant}$ or $\tau_1^{\text{max}} \sim 1/d$.

Equation \eqref{eqn:TransSolBackscatter} is rigorously valid but limited to $\theta = \pi/2$.  We derive an approximate expression for \eqref{eqn:TransSolNasty} that is valid for arbitrary $\theta$ excluding grazing incidence by again using the method of stationary phase.  First, we insert the optical axis coordinates shown in Fig.~\ref{fig:2}, which along the exit surface yield the replacements 
$x = (x_\text{o}+d\cos\theta)/\sin\theta$ and $2\zeta/\sin\theta = c\tau + x_\text{o}/\tan\theta$.  Next, we again assume that $\abs{\chi_0} \ll \sin^4\theta$, and that the transverse spot size is sufficiently large so that the Rayleigh range associated with the projected beam size is much shorter than the propagation length through the crystal, $k_0\sigma_x^2\sin^2\theta \gg (d/\sin\theta + c\tau)$.  Finally, we assume that the transverse size of the beam projected onto $z$ is much smaller than the crystal thickness, so that $\abs{\sigma_x\cos\theta} \ll d$.  These conditions typically apply for all but small grazing incidence angles $\theta \ll 1$, and when satisfied result in a stationary point that is identical to that of Bragg reflection \eqref{eqn:stationaryPt}: $x_s' = (x_\text{o} - c\tau\cot \theta)/\sin\theta$.  Thus, the transmitted field amplitude is
\be
\begin{split}
 E_0\big{\rvert}_P &= -\frac{d \pi^2}{\Lambda^2\sin\theta}
	e^{i\chi_0 k_0(d + c\tau/\sin\theta)/2\sin\theta}  \\
  &\phantom{=|-} \times \exp\!\left\{ -\frac{1}{4\sigma_x^2}\left[(x_\text{o} - c\tau\cot \theta)^2\right] \right\}  \\
  &\phantom{=|-} \times \frac{J_1\!\left[\pi\sqrt{c\tau(2d/\sin\theta + c\tau/\sin^2\theta)}/\Lambda\right]}
  	{\pi\sqrt{c\tau(2d/\sin\theta + c\tau/\sin^2\theta)}/\Lambda}. 	\label{eqn:TransSolApprox}
\end{split}
\ee
From \eqref{eqn:TransSolApprox}, we see that the longitudinal time profile is nearly the same as that for exact backscattering \eqref{eqn:TransSolBackscatter} with $d \rightarrow d/\sin\theta$ [this identification is exact for delays $\tau \ll (d/c)\sin\theta$ if we assume that the Rayleigh range is sufficiently long].  Additionally, we recall that the extinction length $\Lambda$ is nearly independent of $\theta$, so that the temporal envelope $\sim J_1(y)/y$ can be obtained from that evaluated at $\theta = \pi/2$ Eq.~\eqref{eqn:TransSolBackscatter} by replacing $\tau \rightarrow \tau/\sin\theta$.  Thus, the maxima at arbitrary $\theta$ can be written in terms of those at $\theta=\pi/2$ Eq.~\eqref{eqn:maxima0} via
\be
  \tau^{\text{max}}_n(\theta) = \tau^{\text{max}}_n(\pi/2)\sin\theta	\label{eqn:TmaxGeneral}
\ee
with, when $d \gg \mathscr{J}_n \Lambda/\pi$,
\be
  \tau^{\text{max}}_n(\pi/2) = \frac{\mathscr{J}_n^2}{2\pi^2} \frac{\Lambda^2}{cd}.	\label{eqn:TmaxPiH}
\ee

We show in Fig.~\ref{fig:4}(c) the time response of Bragg forward diffraction at three different incidence angles from the (004) Bragg reflection in diamond.  The crystal thickness has been fixed at $d=0.15$ mm, and we plot the power profiles as a function of $t/\sin\theta$.  The position of the intensity maxima is invariant on the normalized time scale, while the power scales as $(d/\sin\theta)^2$.

\begin{figure*}
\centering
\includegraphics[scale=.95]{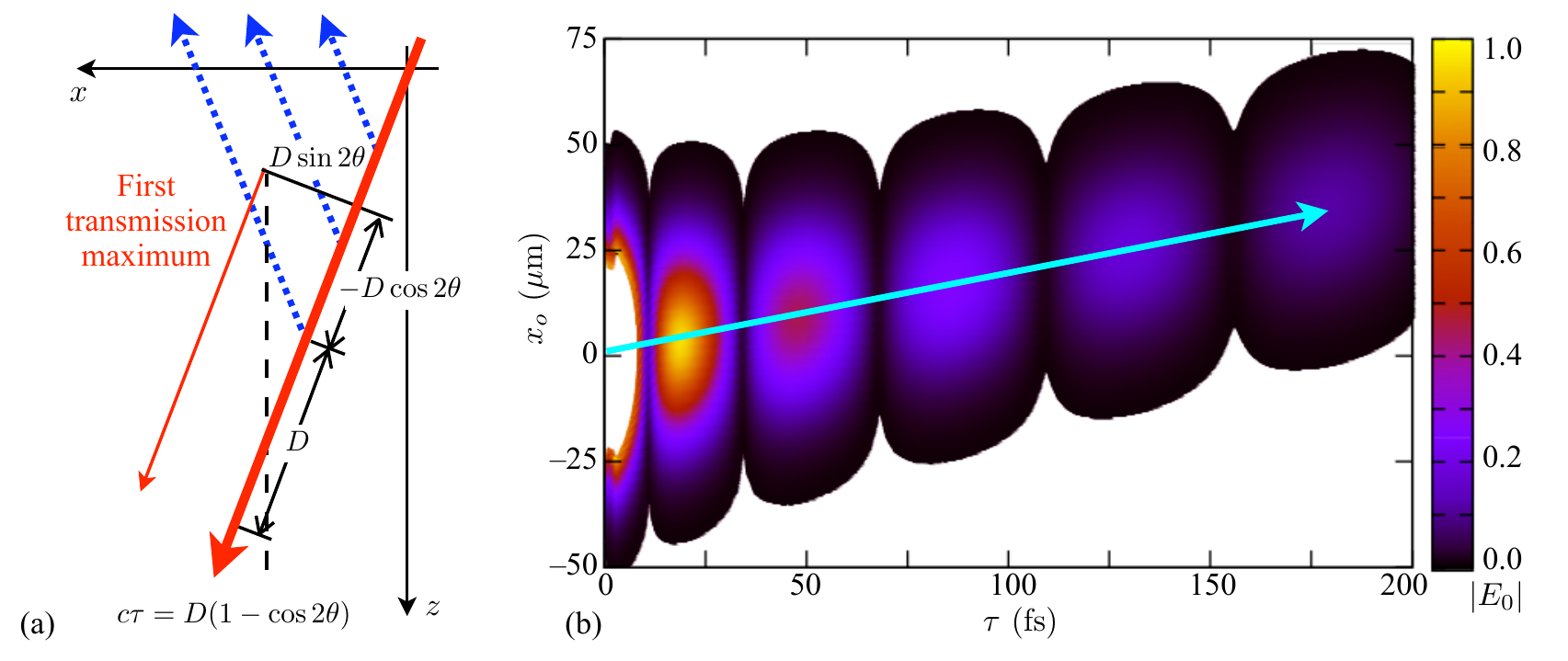}
\caption{(a) Illustration of the coupled forward and reflected modes in the crystal.  As explained in the text, a forward scattered ray is displaced $\Delta x = D\sin 2\theta$ laterally while delayed by the time $\tau = D(1-\cos2\theta)/c$; eliminating $D$, we find that $\Delta x = c\tau \cot\theta$.  (b) Spatiotemporal dependence of the electric field magnitude in forward Bragg diffraction.  We chose the (004) reflection from a 0.1-mm-thick diamond crystal at an incidence angle $\theta = 56.86^\circ$ ($\lambda_0 = 1.5$ \AA), while the incident pulse has an rms width $\sigma_x = 10 \; \mu\text{m}$.  The temporal dependence can be compared with that plotted in Fig.~\ref{fig:4}(c), while the lateral displacement $\Delta x_o = c\tau\cot \theta$ is emphasized by the cyan arrow.}	\label{fig:5}
\end{figure*}

The temporal profile is modified as indicated in \eqref{eqn:TmaxGeneral}, while \eqref{eqn:TransSolApprox} also shows that the transverse envelope is translated from the optical axis at $x_\text{o}=0$ by the amount $c\tau \cot \theta$.  This shift can be related to the distance along $x_0$ that the reflected wave travels during the time $\tau$.  We show this geometrically in Fig.~\ref{fig:5}(a): much like how the delay arises because of the induced Bragg scattering in the crystal, the representative rays are now directed along both $z$ and $x$, so that as energy oscillates between waves the field is displaced in $x_\text{o}$.  Here, the relevant shift can be found by comparing the times $\tau$ and transverse coordinates $x_\text{o}$ after the rays propagate along the characteristics some fixed distance.  From Fig.~\ref{fig:5}(a), we see that while the incident ray propagates a distance $D$, the reflected wave is displaced in time by an amount $c\tau = D - D\cos 2\theta$ and transversely by $x_\text{o} = D\sin 2\theta$.  Taking the ratio, we find that corresponding to a time delay $\tau$, we have the displacement $x_\text{o} = c\tau\sin 2\theta/(1-\cos 2\theta) = c\tau\cot\theta$.

We show the spatiotemporal profile of a Bragg forward scattered pulse from the C(004) crystal at $\theta = 56.86^\circ$ in Fig.~\ref{fig:5}(b).  We assume a temporally short input pulse with rms transverse width $\sigma_x = 10 \; \mu\text{m}$.  Again, we plot the magnitude $\abs{E_0}$ to more clearly show all the trailing pulses, and we scale the amplitude so that the first delayed maxima has unit magnitude.  The temporal profile closely mirrors that shown of the same crystal in Fig.~\ref{fig:4}(c), while each subsequent pulse is more distantly displaced from the optical axis $x_{\text{o}} = 0$.  The cyan line is drawn along the theoretically predicted line $x_{\text{o}} = c\tau \cot \theta$, which closely predicts the transverse displacement.  In this case, the first trailing pulse is displaced from the optical axis by a small amount $\sim 5 \; \mu$m, while the second maxima is shifted by an amount $\sim 13 \; \mu$m, which is of the order of the rms width.

\section{Monochromatic power for FEL self-seeding}

We have seen that Bragg forward scattering gives rise to a sequence of delayed power maxima from temporally short incident radiation.  Reference \cite{Geloni_etal:2011} proposed using the first trailing maximum to seed an FEL at hard x-ray wavelengths: an initially short SASE pulse generates the radiation ``wake'' depicted in Fig.~\ref{fig:5}(b) that is then used to coherently seed the FEL interaction in downstream undulators.  In the preceding discussion, we computed the spatiotemporal field profile generated by an initially short and coherent incident pulse; since SASE is temporally incoherent (chaotic light), however, some additional considerations are necessary to determine the relevant radiation seed power.

The longitudinal structure of SASE can be well modeled as a sum of Gaussian modes that have random temporal positions and phases.  For $M$ longitudinal modes, we approximate the SASE field as
\be
  E_{\text{inc}}(t) = U(t) \sum_{j=1}^M \frac{\epsilon_j}{\sqrt{4\pi} \sigma_\tau}
  	e^{-c^2(t-t_j)^2/4\sigma_\tau^2},	\label{eqn:inciSASE}
\ee
where $U(t)$ is the total envelope of the field determined by the electron beam current and FEL gain, $t_j$ are a random set of times, and $\epsilon_j$ are a set of random complex amplitudes such that $M\langle \abs{\epsilon_j}^2 \rangle/\sigma_\tau$ is proportional to the ensemble-averaged SASE energy.  The temporal width $\sigma_\tau$ is dictated by the FEL physics while the number of modes $M$ is determined by $\sigma_\tau$ and the characteristic length of the envelope $U(t)$ that we denote $L_{\text{pulse}}$.  Assuming that $\sigma_\tau$ is sufficiently short, the incident pulse \eqref{eqn:inciSASE} generates a trailing electromagnetic field that is merely a sum of $M$ wakes discussed previously, with each one beginning at the time $t_j$ and having the relative complex amplitude given by $\epsilon_j$.

First, we consider the case when the duration of the incident SASE length $L_{\text{pulse}}$ is much shorter than the characteristic time-scale of the trailing wake, which is the typical situation for self-seeding of few-fs pulses.   In this case, the $M$ Bragg forward-diffracted beams generated by \eqref{eqn:inciSASE} have the same temporal shape, but differ with random phases and amplitudes.  The ensemble-averaged total power in the trailing pulse scales as $M$ times the average power in an individual wake.  If we consider the power maxima of \eqref{eqn:TransSolApprox}, the ratio of the ensemble-averaged power in the first trailing seed to that in the SASE is
\begin{align}
  \frac{P_{\text{seed}}^{\text{max}}}{P_{\text{pulse}}} &= 4\pi \sigma_\tau^2 M 
  	\left[\frac{\pi^2 d}{\Lambda^2\sin\theta}\frac{J_1(\mathscr{J}_1)}{\mathscr{J}_1}\right]^2,
		\label{eqn:Pratio}	
\end{align}
where again $\mathscr{J}_1$ is the position of the first maxima of $\abs{J_1(y)/y}$ with $y>0$.  The temporal width of each mode and number of modes is dictated by the FEL process, and as such will depend on the electron beam quality and profile.  In order to get an approximate expression, we use the analytic results that are available for a long electron beam with zero initial energy spread \cite{Kim:1986,
WangYu:1986}.  In this case, the rms width of a SASE mode is related to the FEL coherence length $L_{\text{coh}}$ by $\sigma_\tau = L_{\text{coh}}/\sqrt{3\pi}$ while the number of longitudinal modes for a pulse of total length $L_{\text{pulse}}$ is $M = L_{\text{pulse}}/L_{\text{coh}}$; inserting this into \eqref{eqn:Pratio} yields
\be
  \frac{P_{\text{seed}}^{\text{max}}}{P_{\text{pulse}}} \sim \frac{4}{3} L_{\text{coh}} L_{\text{pulse}}
  	\left[\frac{\pi^2 d}{\Lambda^2\sin\theta} \frac{J_1(\mathscr{J}_1)}{\mathscr{J}_1} \right]^2.
\ee
Finally, the SASE coherence length generated by an electron beam of vanishing energy spread is
\be
  L_{\text{coh}} = \frac{\lambda_0}{6\rho} \sqrt{\frac{N_G}{2\pi}},	\label{eqn:LcohSASE}
\ee
where $N_G$ is the number of gain lengths in the upstream undulator.  Combining \eqref{eqn:LcohSASE} and \eqref{eqn:Pratio}, we have
\begin{align}
  \frac{P_{\text{seed}}^{\text{max}}}{P_{\text{pulse}}} &\sim \sqrt{\frac{2N_G}{\pi}} 
  	\frac{\lambda_0 L_{\text{pulse}}}{9\rho} \left[\frac{\pi^2 d}{\Lambda^2\sin\theta}
	\frac{J_1(\mathscr{J}_1)}{\mathscr{J}_1} \right]^2	\nonumber  \\
  &\approx 0.0058 \sqrt{\frac{N_G}{2\pi}} \frac{\lambda_0 L_{\text{pulse}}}{6\rho}
  	\left[\frac{\pi^2 d}{\Lambda^2\sin\theta}\right]^2.
  	\label{eqn:PratioFin}	
\end{align}
Due to the various approximations made, we only expect \eqref{eqn:PratioFin} to be correct to within a factor of two or so.  Nevertheless, this yields a useful estimate and has the appropriate scaling when $L_{\text{pulse}}$ is much less than the time scale of variation of the forward Bragg diffraction signal; if we use \eqref{eqn:TmaxGeneral}-\eqref{eqn:TmaxPiH} to eliminate the crystal parameters in favor of the delay, \eqref{eqn:PratioFin} becomes
\be
\frac{P_{\text{seed}}^{\text{max}}}{P_{\text{pulse}}} \approx  \sqrt{\frac{N_G}{2\pi}} 
	\frac{\lambda_0 L_{\text{pulse}}}{6\rho} \frac{1}{(c\tau_1)^2} =
	\frac{L_{\text{coh}}L_{\text{pulse}}}{(c\tau_1)^2}.	\label{eqn:PratioAlt}
\ee
For example, for the approximate parameters proposed in \cite{Geloni_etal:2010b} for use at the Linac Coherent Light Source (LCLS), the low charge operation has $L_{\text{pulse}} \approx 1 \; \mu\text{m}$ and after the suggested interaction length we find that $L_{\text{coh}} \approx 0.07 \; \mu\text{m}$; for a 0.1-mm diamond crystal at an angle $\theta = 56.86^\circ$, the C(004) planes have $\pi^2 d/(\Lambda^2\sin\theta) \approx 2.4 \; \mu\text{m}^{-1}$, and \eqref{eqn:PratioFin} implies that a 1-GW SASE pulse would produce a seed whose ensemble-averaged peak power is 2.3 MW; the same result obtains from \eqref{eqn:PratioAlt} for the associated delay $\tau_1 \approx 18$ fs.  These estimates compare quite favorably with the quoted result of about 2.5 MW \cite{Geloni_etal:2010b} .

\begin{figure}
\centering
\includegraphics{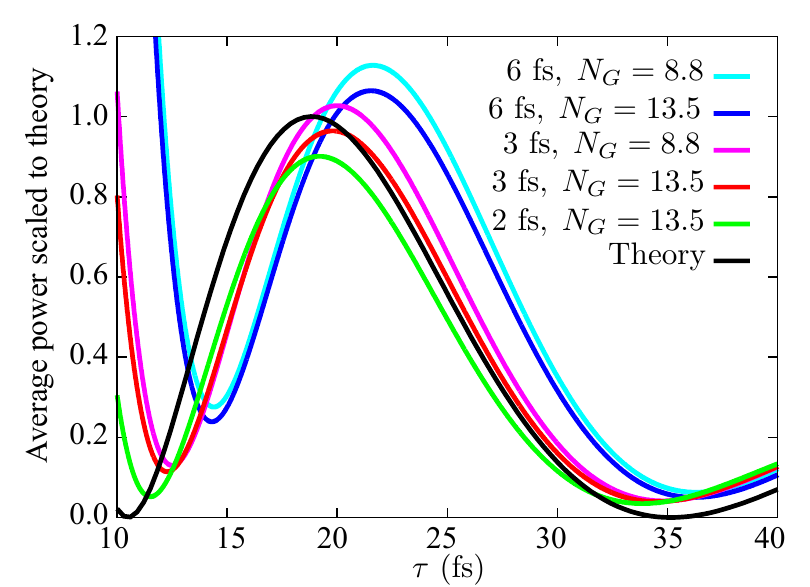}
\caption{Ensemble-averaged power of the monochromatic seed obtained from simulation, scaled to the theoretical predicted amplitude \eqref{eqn:PratioFin}.  We have varied both the electron beam temporal duration and the undulator length to produce the $L_\text{pulse}$ and $N_G$ listed in the key.  We have used a flat-top electron beam profile to more accurately determine $L_{\text{pulse}}$; more realistic distributions can be expected to yield power deviations from the theory by a factor of two.  The position of the first maximum linearly increases as the pulse duration increases, and therefore varies by a few femtoseconds.}	\label{fig:6}
\end{figure}

To demonstrate the scaling predicted in \eqref{eqn:PratioFin} holds in general (assuming that $L_{\text{pulse}} \ll \Lambda^2/\pi^2 d$), we used a simple 1D FEL code to generate SASE output for several different pulse lengths $L_{\text{pulse}}$ and over two different number of gain lengths to compute the seeding wake after the monochromator for LCLS-type parameters cited above.  We show the results in Fig.~\ref{fig:6}, where we scale the seed power with the ratio \eqref{eqn:PratioFin}, so if our results are exact all the lines should overlap with the theory line.  We note that the seed power is correctly predicted within 10\% or so, while the delay of the seed increases over that predicted by the theory by an amount approximately equal to $L_{\text{pulse}}$.  For short pulses, this shift is a minor correction.

The power statistics of the seeding wake are identical to that of a single SASE intensity mode/spike, meaning that the effective seed power fluctuates 100\%.  In the frequency domain, this reflects the fact that the power fluctuations of chaotic light within a bandwidth much less than $c/\sigma_\tau$ approach unity.

Finally, we wish to make a few statements in the application of Bragg forward scattering to the seeding of relatively long pulses.  By long pulses, we mean those for which $L_{\text{pulse}} \gtrsim \Lambda^2\sin\theta/\pi^2 d$; for the LCLS parameters used herein, this applies for pulses longer than about 10 fs.  In this case, the temporal modes in \eqref{eqn:inciSASE} are spread over a time that is longer than that associated with Bragg forward scattering.  Thus, we expect that the effective number of modes contributing to the seed amplitude to be $M \lesssim (\Lambda^2\sin\theta/\pi^2 d)/\sigma_\tau$.  While the peak power in this trailing monochromatic seed is therefore of order that in the short pulse case, one must confront the fact that as the delay $\tau$ is increased to accommodate the longer pulse, the wake is displaced transversely according to $x_\text{o} = c\tau\cot\theta$.  Since a typical FEL-produced radiation beam has $\sigma_x \sim 20 \; \mu\text{m}$ and $\cot\theta \sim 1$ to allow variations in the central wavelength, this limits the applicability of this self-seeding scheme to $L_{\text{pulse}}/c \lesssim 50$ fs.

\section{Conclusions and future directions}

We have used the dynamical theory of x-ray diffraction to calculate approximate, analytic expressions for both the reflected and forward scattered electric field amplitudes resulting from Bragg diffraction of an initially short x-ray pulse incident on a crystal.  Both field profiles are characterized by a temporal envelope that oscillates in time $\sim J_1(y)/y$, where $y \propto \tau/\sin\theta$ for the reflected wave while $y \propto \sqrt{d\tau/\sin\theta}$ for the forward diffracted field. The delayed output is also displaced transversely by an amount $x_o = c\tau \cot\theta$,  which can be associated with the coupled wave interaction in the crystal.  Finally, we used the developed theory to analyze in detail the dynamics of the ``wake monochromator'' as applied to self-seeding for hard x-ray free-electron lasers.  We found simple relationship for the induced delay and output power in terms of the crystal and SASE parameters.

In the future we plan to extend this treatment to asymmetrically cut crystals in both Bragg and Laue scattering geometries, which can be used to analyze the field output generated by temporally short x-ray pulses from a wide variety of x-ray optical elements.

\acknowledgments
The authors would like to thank A.~A.~Zholents in particular and the LCLS Hard X-Ray Self-Seeding collaboration in general for stimulating interest in this topic and K.-J.~Kim for useful discussions.  Work supported by U.S. Dept.~of Energy Office of Sciences under Contract No.~DE-AC02-06CH11357.

\appendix

\section{Riemann's method and application to Bragg scattering}

Any linear second-order differential operator $\mathcal{L}$ that is also hyperbolic can be written as acting on the function $\mathcal{A}$ via (see, e.g., \cite{garabedian:1998})
\be
  \mathcal{L}(\mathcal{A}) \equiv \frac{\partial^2}{\partial \zeta' \partial\xi'}\mathcal{A}
  	+ a\frac{\partial}{\partial\zeta'}\mathcal{A} + b\frac{\partial}{\partial \xi'}\mathcal{A} + c\mathcal{A},
\ee
where $(\zeta', \xi')$ are the characteristic coordinates.  We additionally consider the adjoint $\mathcal{L}^\dagger$ of the operator $\mathcal{L}$, which is defined through the inner product relationship $\int R\mathcal{L}(\mathcal{A}) = \int \mathcal{L}^\dagger(R)\mathcal{A}$.  This definition implies that $R\mathcal{L}(\mathcal{A}) - \mathcal{L}^\dagger(R)\mathcal{A}$ can be written as the divergence of a vector function
\be
  R \, \mathcal{L}(\mathcal{A}) - \mathcal{L}^\dagger(R)\, \mathcal{A} = \bv{\nabla}\cdot\bv{P} = 
  	\frac{\partial P_{\zeta'}}{\partial\zeta'} + \frac{\partial P_{\xi'}}{\partial\xi'} 	\label{eqn:diffAdjointDef}
\ee
with $\bv{P}$ vanishing at the endpoints.  Thus, $\mathcal{L}^\dagger$ is completely specified by both the equation \eqref{eqn:diffAdjointDef} and by a set of boundary conditions.  Ignoring for the moment these boundary conditions, one can easily show that the differential identity \eqref{eqn:diffAdjointDef} is satisfied if the action of adjoint $\mathcal{L}^\dagger$ is given by \cite{garabedian:1998}
\be
  \mathcal{L}^\dagger(R) \equiv \frac{\partial^2}{\partial \zeta' \partial\xi'}R + \frac{\partial}{\partial\zeta'}(aR)
  	+ \frac{\partial}{\partial \xi'}(bR) + cR = 0,
\ee
and the components of $\bv{P} = (P_{\zeta'}, P_{\xi'})$ are
\begin{align}
  P_{\zeta'} &= \frac{1}{2}\left(R\frac{\partial \mathcal{A}}{\partial\zeta'} -
  	\mathcal{A}\frac{\partial R}{\partial\zeta'}\right) + aR\mathcal{A}  \\
  P_{\xi'} &= \frac{1}{2}\left(R\frac{\partial \mathcal{A}}{\partial\xi'} -
  	\mathcal{A}\frac{\partial R}{\partial\xi'}\right) + bR\mathcal{A}.
\end{align}
Riemann's method is obtained by integrating the differential identity \eqref{eqn:diffAdjointDef} over an arbitrary closed region $\Gamma$ in the $(\zeta',\xi')$ plane.  Using Green's identity on the right-hand side, we find
\be
  \int\limits_\Gamma \! d\sigma \; [R \mathcal{L}(\mathcal{A}) - \mathcal{A} \mathcal{L}^\dagger(R)] = 
  	\oint\limits_\gamma \! d\ell \;  \hat{\gamma}\cdot \bv{P},	\label{eqn:RiemannsEqn}
\ee
where the integration proceeds along the boundary $\gamma$ and $\hat{\gamma}$ is its outward-facing normal.  Now, if we assume that $\mathcal{A}$ is a solution to $\mathcal{L}(\mathcal{A}) = 0$, and $R$ solves the adjoint equation $\mathcal{L}^\dagger(R) = 0$, then $\oint \! d\ell \;  \hat{\gamma}\cdot \bv{P}$ vanishes.  By inserting $\bv{P}$, we have an expression that can be used to find the solution to $\mathcal{L}(\mathcal{A})=0$ (assuming it exists) in terms of its values along the boundary and the Riemann function $R$.

For the problem of Bragg scattering, $a=b=0$, so that $\mathcal{L}$ and $\mathcal{L}^\dagger$ satisfy the same partial differential equation, and $\mathcal{L}$ is formally self-adjoint.  Now, we can find the solution by determining an appropriate curve $\gamma$ and Riemann function $R$ which, due to the differing boundary conditions, will depend on whether we consider forward or backward scattering.  In the latter case, the triangle $ABQ$ is a convenient choice to obtain the reflected wave $\mathcal{A}_h$ at point $Q$, which is chosen such that $\xi'_B \le \xi'_C$ meaning that the solution is given only over the time interval $0 \le t \le 2d/\sin\theta$.  If we further assume that the fields $\mathcal{A}_0$, $\mathcal{A}_h$ vanish along the line $\overline{AB}$, we can integrate \eqref{eqn:RiemannsEqn} by parts to write it as
\be
\begin{split}
  0 = \oint\limits_\gamma \! d\ell \;  \hat{\gamma}\cdot \bv{P} &= R \mathcal{A}_h\big\rvert_Q -
  	\int\limits_B^Q d\zeta' \; \mathcal{A}_h \frac{\partial R}{\partial\zeta'}  \\
  &\phantom{=} - \int\limits_Q^A \frac{d\ell}{\sqrt{2}}\left(\mathcal{A}_h \frac{\partial R}{\partial\xi'}
  	+ R\frac{\partial\mathcal{A}_h}{\partial\zeta'}\right).   \label{eqn:ReflSolAppend}
\end{split}
\ee
Here, the Riemann function $R = R(\zeta, \xi; \zeta', \xi')$ satisfies the adjoint equation associated with the (formally self-adjoint) system \eqref{eqn:ReducedSystem}:
\be
  \frac{\partial^2}{\partial\xi'\partial\zeta'}R(\zeta, \xi; \zeta', \xi') = -\frac{\pi^2}{\Lambda^2}
  	R(\zeta, \xi; \zeta', \xi').	\label{eqn:AdjointAppend}
\ee
If $R$ additionally satisfies $\partial R/\partial\xi' = 0$ along $\overline{AQ}$ (the line $\zeta'=\xi'$) and $\partial R/\partial\zeta' = 0$ along $\overline{BQ}$ (when $\xi'=\xi$), then \eqref{eqn:ReflSolAppend} gives the solution for the reflected wave in terms of the incident wave $\mathcal{A}_0 \propto \partial\mathcal{A}_h/\partial\zeta'$.  The Riemann function that satisfies these requirements and also $R(Q) = R(\zeta,\xi;\zeta',\xi') = 1$ is \cite{AfanasevKohn:1971}
\be
\begin{split}
  R_h(\zeta, \xi; \zeta', \xi') = J_0\!\left[2\pi\sqrt{(\zeta-\zeta')(\xi-\xi')}/\Lambda\right]  \hspace{.25 in} \\
  + \frac{\xi-\xi'}{\zeta-\zeta'}J_2\!\left[2\pi\sqrt{(\zeta-\zeta')(\xi-\xi')}/\Lambda\right].
  		\label{eqn:RiemannReflAppend}
\end{split}
\ee
For the Riemann function $R_h$ only the integration along $\overline{AQ}$ contributes from \eqref{eqn:ReflSolAppend}, and the field is therefore given by
\be
  \mathcal{A}_h\big\rvert_Q = \int\limits_A^Q \! d\zeta' \; \frac{ik_0 \chi_{\vphantom{\bar{h}}h}}{2\sin\theta}
  	\mathcal{A}_0(\zeta',\zeta') R_h(\zeta, \xi; \zeta', \zeta').	\label{eqn:Refl_IntSolAppend}
\ee

To determine the equation for the transmitted wave at point $P$ on the rear surface of the crystal $\overline{CP}$, we choose the contour $\gamma$ to be the parallelogram $ACPQ$, which is restricted to $d/\sin\theta \le t \le 3d/\sin\theta$ to exclude the effects of surface reflections.  Again assuming that the fields $\mathcal{A}_0$ and $\mathcal{A}_h$ vanish along the line $\overline{AB}$, \eqref{eqn:RiemannsEqn} can be partially integrated to yield
\be
\begin{split}
  0 = \oint\limits_\gamma \! d\ell \;  \hat{\gamma}\cdot \bv{P} &= R \mathcal{A}_0\big\rvert_P^Q +
  	\int\limits_P^Q d\xi' \; \mathcal{A}_0 \frac{\partial R}{\partial\xi'}  \\
  &\phantom{=} + \int\limits_Q^A \frac{d\ell}{\sqrt{2}}\left(\mathcal{A}_0 \frac{\partial R}{\partial\xi'}
  	+ R\frac{\partial\mathcal{A}_0}{\partial\zeta'}\right)  \\
  &\phantom{=} - \int\limits_C^P \frac{d\ell}{\sqrt{2}}\left(R \frac{\partial \mathcal{A}_0}{\partial\xi'}
  	+ \mathcal{A}_0\frac{\partial R}{\partial\zeta'}\right).   \label{eqn:TransSolAppend}
\end{split}
\ee
Along the front surface the incident wave $\mathcal{A}_0$ is prescribed while the reflected wave $\mathcal{A}_h$ is given by \eqref{eqn:Refl_IntSolAppend}.  On the other hand, the reflected wave $\mathcal{A}_h \propto \partial\mathcal{A}_0/\partial\xi'$ vanishes along the rear surface $\overline{CP}$.  Thus, to reduce \eqref{eqn:TransSolAppend} to a closed form solution, we require that the Riemann function $R_0$ solving \eqref{eqn:AdjointAppend} also satisfies the auxilliary conditions
\begin{align}
  \left. \frac{\partial R_0}{\partial\xi'}\right\rvert_{\overline{PQ}} &= 0,  &
  	\left. \frac{\partial R_0}{\partial\zeta'}\right\rvert_{\overline{CP}} &= 0,  &  R_0(P) &= 1.
		\label{eqn:TransAux}
\end{align}
The Riemann function satisfying the boundary conditions \eqref{eqn:TransAux} and the adjoint equation \eqref{eqn:AdjointAppend} associated with the transmitted wave is \cite{AfanasevKohn:1971}
\be
\begin{split}
  R_0(\zeta, \xi; \zeta', \xi') = J_0\!\left[2\pi \sqrt{(\zeta-\zeta')(\xi-\xi')}/\Lambda \right]  \hspace{.25 in} \\
  + \frac{\zeta-\zeta'}{\xi-\xi'}J_2\!\left[2\pi \sqrt{(\zeta-\zeta')(\xi-\xi')}/\Lambda \right],
\end{split}
\ee
while rearranging \eqref{eqn:TransSolAppend} implies that the solution
\beu
\begin{split}
  \mathcal{A}_0\big\rvert_P &= R_0 \mathcal{A}_0\big\rvert_Q - \int\limits_A^Q \! d\zeta' \;
	R_0(\zeta,\xi;\zeta',\zeta')\frac{ik_0\chi_{\bar{h}}}{2\sin\theta}\mathcal{A}_h  \\
  &\phantom{=} - \int\limits_A^Q \! d\zeta' \, \left.\mathcal{A}_0(\zeta',\zeta')\frac{\partial}{\partial\zeta'}
  	R_0(\zeta,\xi;\xi',\zeta')\right\rvert_{\xi'=\zeta'}.
\end{split}
\eeu

\end{document}